\documentclass[aps,showpacs,byrevtex,twocolumn]{revtex4}
\usepackage{amsmath}
\usepackage{amssymb}
\usepackage[dvips]{graphicx}
\usepackage{latexsym}

\newcommand{\real}{{\mathcal R}e}
\newcommand{\imag}{{\mathcal I}m}

\begin{document}

\title{Numerical approximations using
       Chebyshev polynomial expansions: El-gendi's method revisited}

\author{Bogdan Mihaila}
\email{bogdan.mihaila@unh.edu} \affiliation{Physics Division,
Argonne National Laboratory, Argonne, IL 60439}

\author{Ioana Mihaila}
\email{mihaila@mailbox.coastal.edu} \affiliation{Department of
Mathematics and Statistics, Coastal Carolina University,P.O. Box
261954, Conway, SC 29528-6054}

\begin{abstract}
We present numerical solutions for differential equations by
expanding the unknown function in terms of Chebyshev polynomials
and solving a system of linear equations directly for the values
of the function at the extrema (or zeros) of the Chebyshev
polynomial of order~$N$ (El-gendi's method). The solutions are
exact at these points, apart from round-off computer errors and
the convergence of other numerical methods used in solving the
linear system of equations. Applications to initial value problems
in time-dependent quantum field theory, and second-order boundary
value problems in fluid dynamics are presented.
\end{abstract}


\maketitle

\section{Introduction}

A major problem in modern physics is to understand the time
evolution of a quark-gluon plasma produced following a
relativistic heavy-ion collision. Although a mean field
theoretical approach~\cite{ref:Hartree,ref:GV,ref:LOLN} can
provide a reasonable picture of the phase diagram of quantum field
theories, such studies do not include a rescattering mechanism,
which would allow an out of equilibrium system to be driven back
to equilibrium. As such, the past few years have witnessed a major
effort concerning the search for approximation
schemes~\cite{ref:EQT,ref:abw,ref:ctpN,ref:berges} which go beyond
mean field theory. In the process, new numerical techniques were
required in order to solve the ever more challenging systems of
complex integro-differential equations.

In this paper we revive and extend an old spectral method based on
expanding the unknown function in terms of Chebyshev polynomials,
which plays a crucial role in implementing our non-equilibrium
field theory program. Finite-difference methods, though leading to
sparse matrices, are notoriously slowly convergent. Thus the need
to use higher-order methods, like the nonuniform-grid Chebyshev
polynomial methods, which belong to a class of spectral numerical
methods. Then the resulting matrices are less sparse, but what is
apparently lost in storage requirements, is regained in speed. We
do in fact keep the storage needs moderate, as we can achieve very
good accuracy with a moderate number of grid points.

The Chebyshev polynomials of the first kind of degree~$n$,
$T_n(x)$ with $n\le N$, satisfy discrete orthogonality
relationships on the grid of the 
extrema of $T_N(x)$. Based on this property, Clenshaw and
Curtis~\cite{ref:Clenshaw
} proposed almost forty years ago a quadrature scheme for finding
the integral of a non-singular function defined on a finite range,
by expanding the integrand in a series of Chebyshev polynomials
and integrating this series term by term. Bounds for the errors of
the quadrature
scheme have been discussed in \cite{ref:errors1
} and reveal that by truncating the series at some order~$m < N$
the difference between the exact expansion and the truncated
series can not be bigger than the sum of the neglected expansion
coefficients~\cite{ref:nr}. 
This is a consequence of the fact that the Chebyshev polynomials
are bounded between $\pm 1$, and if the expansion coefficients are
rapidly decreasing, then the error is dominated by 
the $m+1$ term of the series, and spreads out smoothly over the
interval $[-1,1]$.

Based on the discrete orthogonality relationships of the Chebyshev
polynomials, various methods for solving linear and nonlinear
ordinary differential equations \cite{ref:ode1
} and integral differential equations \cite{ref:inteqn} were
devised at about the same time and were found to have considerable
advantage over finite-difference methods. Since then, these
methods have become standard and are part of the larger family of
spectral methods~
\cite{ref:Boyd
}.
They rely on expanding out the unknown function in a large series
of Chebyshev polynomials, truncating this series, substituting the
approximation in the actual equation, and determining equations
for the coefficients. El-gendi~\cite{ref:El-gendi} has argued
however, that it is better to compute directly the values of the
functions rather than the Chebyshev coefficients. The two
approaches are formally equivalent in the sense that if we have
the values of the function, the Chebyshev coefficients can be
calculated.

In this paper we use the discrete orthogonality relationships of
the Chebyshev polynomials to discretize various continuous
equations by reducing the study of the solutions to the Hilbert
space of functions defined on the set of (N+1) extrema of
$T_N(x)$, spanned by a discrete (N+1)-term Chebyshev polynomial
basis.
In our approach we follow closely the procedures outlined by
El-gendi~\cite{ref:El-gendi} for the calculation of integrals, but
extend his work to the calculation of derivatives. We also show
that similar procedures can be applied for a second grid given by
the zeros of $T_N(x)$.

In our presentation we shall leave out the technical details of
the physics problems, and shall refer the reader to the original
literature instead. Also, even though our main interest regards
the implementation of the Chebyshev method for solving initial
value problems, we present a perturbative approach for solving
boundary value problems, which may be of interest for fluid
dynamics applications.

The paper is organized as follows: In
Section~\ref{sec:Fundamentals} we review the basic properties of
the Chebyshev polynomial and derive the general theoretical
ingredients that allow us to discretize the various equations. The
key element is the calculation of derivatives and integrals
without explicitly calculating the Chebyshev expansion
coefficients. In Sections~\ref{sec:IVP} and~\ref{sec:BVP} we apply
the formalism to obtain numerical solutions of initial value and
boundary value problems, respectively. We accompany the general
presentation with examples, and compare the solution obtained
using the proposed Chebyshev method with the numerical solution
obtained using the finite-difference method. Our conclusions are
presented in Section~\ref{sec:Conclusions}.

\section{Method of Chebyshev expansion}
\label{sec:Fundamentals}

The Chebyshev polynomial of the first kind of degree~$n$, $T_n(x)$,
has $n$ zeros in the interval $[-1, 1]$, which are located at the
points
\begin{equation}
   x_k \ = \
   \cos \left ( \frac{\pi (k - \frac{1}{2})}{n} \right ) \>,
   \quad k = 1,2,\ldots,n
   \>.
   \label{eq:Tn_zeros}
\end{equation}
In the same interval the polynomial $T_n(x)$ has $n+1$ extrema
located at
\begin{equation}
   \tilde{x}_k \ = \
   \cos \left ( \frac{\pi k}{n} \right ) \>,
   \quad k = 0,1,\ldots,n
   \>.
   \label{eq:Tn_max}
\end{equation}
The Chebyshev polynomials are orthogonal in the interval $[-1,
1]$ over a weight~$(1-x^2)^{-1/2}$. In addition, the Chebyshev
polynomials also satisfy discrete orthogonality relationships.
These correspond to the following choices of grids:
\begin{itemize}
   \item    If $x_k \ (k = 1,2,\ldots,N)$ are the {\em N} zeros of $T_N(x)$
            given by (\ref{eq:Tn_zeros}), and if $i,j < N$, then
            \begin{equation}
               \sum_{k=1}^N \ T_i(x_k) T_j(x_k)
               \ = \
               \alpha_i \ \delta_{i \, j}
               \>,
            \label{eq:cheby_ortog_a}
            \end{equation}
            where the constants $\alpha_i$ are
            \[
               \alpha_i
               \ = \
               \left \lbrace
                  \begin{array}{ll}
                     \displaystyle{\frac{N}{2}}  \>, & i \neq 0 \>, \\
                     N                           \>, & i = 0 \>.
                  \end{array}
               \right .
            \]
    \item   If $\tilde{x}_k$ are defined by~(\ref{eq:Tn_max}),
            then the discrete orthogonality relation is
            \begin{equation}
               \sum_{k=0}^N {\rm {}''} \ T_i(\tilde{x}_k) T_j(\tilde{x}_k)
               \ = \
               \beta_i \ \delta_{i \, j}
               \>,
            \label{eq:cheby_ortog_b}
            \end{equation}
            where the constants $\beta_i$ are
            \[
               \beta_i
               \ = \
               \left \lbrace
                     \begin{array}{ll}
                        \displaystyle{\frac{N}{2}}  \>, & i \neq 0,N \>,\\
                        N                           \>, & i = 0,N \>.
                     \end{array}
               \right .
            \]
\end{itemize}
Here, the summation symbol with double primes denotes a sum
with both the first and last terms halved.


In general, we shall seek to approximate the values of the
function $f$ corresponding to a given discrete set of points like
those given in Eqs.~(\ref{eq:Tn_zeros}, \ref{eq:Tn_max}). Using
the orthogonality relationships, Eqs.~(\ref{eq:cheby_ortog_a},
\ref{eq:cheby_ortog_b}), we have a procedure for finding the
values of the unknown function (and any derivatives or
anti-derivatives of it) at either the zeros or the local extrema
of the Chebyshev polynomial of order~$N$.

A continuous and bounded variation function $f(x)$ can be
approximated in the interval $[-1,1]$ by either one of the two
formulae
\begin{eqnarray}
   f(x)
   & \approx &
   \sum_{j=0}^{N-1} {\rm {}'} \ a_j T_j(x)
   \>,
\label{eq:f_approx_a}
\end{eqnarray}
or
\begin{eqnarray}
   f(x)
   & \approx &
   \sum_{j=0}^{N} {\rm {}''} \ b_j \ T_j(x)
   \>,
\label{eq:f_approx_b}
\end{eqnarray}
where the coefficients $a_j$ and $b_j$ are defined as
\begin{eqnarray}
   a_j
   & = &
   \frac{2}{N} \,
   \sum_{k=1}^N \
       f(x_k) T_j(x_k) \>, \quad j = 0,\ldots,N-1
   \>,
\label{eq:coeff_a}
   \\
   b_j
   & = &
   \frac{2}{N} \,
   \sum_{k=0}^N {\rm {}''} \
       f(\tilde{x}_k) T_j(\tilde{x}_k) \>, \quad j = 0,\ldots,N
   \>,
\label{eq:coeff_b}
\end{eqnarray}
and the summation symbol with one prime denotes a sum with the
first term halved. The approximate formulae~(\ref{eq:f_approx_a})
and~(\ref{eq:f_approx_b}) are exact at {\em x} equal to $x_k$
given by Eq.~(\ref{eq:Tn_zeros}), and at $x$ equal to $\tilde x_k$
given by Eq.~(\ref{eq:Tn_max}), respectively.

Derivatives and integrals can be computed at the grid points by
using the expansions (\ref{eq:f_approx_a}, \ref{eq:f_approx_b}).
The derivative $f'(x)$ is approximated as
\begin{eqnarray}
   f'(x)
   & \approx &
   \sum_{k=1}^N \
       f(x_k) \
   \frac{2}{N} \,
   \sum_{j=0}^{N-1} {\rm {}'} \ T_j(x_k) \, T_j'(x)
   \>,
\label{eq:f_derivative_a}
\end{eqnarray}
and
\begin{eqnarray}
   f'(x)
   & \approx &
   \sum_{k=0}^N {\rm {}''} \
       f(\tilde{x}_k)  \
   \frac{2}{N} \,
   \sum_{j=0}^{N} {\rm {}''} \
       T_j(\tilde{x}_k) \ T_j'(x)
   \>.
\label{eq:f_derivative_b}
\end{eqnarray}
Similarly, the integral $\int_{-1}^x f(t) \, {\rm d}t$ can be approximated as
\begin{eqnarray}
   \int_{-1}^{x} f(t) \, {\rm d}t
   & \approx &
   \sum_{k=1}^N \
       f(x_k) \
   \frac{2}{N} \,
   \sum_{j=0}^{N-1} {\rm {}'} \
       T_j(x_k) \, \int_{-1}^{x} T_j(t) \, {\rm d}t
   \>,
 \nonumber \\ &&
\label{eq:f_int_a}
\end{eqnarray}
or
\begin{eqnarray}
   \int_{-1}^{x} f(t) \, {\rm d}t
   & \approx &
   \sum_{k=0}^N {\rm {}''} \
       f(\tilde{x}_k)  \
   \frac{2}{N} \,
   \sum_{j=0}^{N} {\rm {}''} \
       T_j(\tilde{x}_k) \ \int_{-1}^{x} T_j(t) \, {\rm d}t
   \>.
\nonumber \\ &&
\label{eq:f_int_b}
\end{eqnarray}
Thus, one can calculate integrals and derivatives based on the
Chebyshev expansions (\ref{eq:f_approx_a}) and
(\ref{eq:f_approx_b}), avoiding the direct computation of the
Chebyshev coefficients (\ref{eq:coeff_a}) or (\ref{eq:coeff_b}),
respectively. In matrix format we have
\begin{eqnarray}
   \left [ \int_{-1}^x \ f(t) \, {\rm d}t \right ]
   & \approx &
   S \ \left [ f \right ]
   \>,
\label{eq:Beqn_a}
   \\
   \left [ f'(x) \right ]
   & \approx &
   D \ \left [ f \right ]
   \>,
\label{eq:BTeqn_a}
\end{eqnarray}
for the case of the grid (\ref{eq:Tn_zeros}), and
\begin{eqnarray}
   \left [ \int_{-1}^x \ f(t) \, {\rm d}t \right ]
   & \approx &
   \tilde S \ \left [ f \right ]
   \>,
\label{eq:Beqn_b}
   \\
   \left [ f'(x) \right ]
   & \approx &
   \tilde D \ \left [ f \right ]
   \>,
\label{eq:BTeqn_b}
\end{eqnarray}
for the case of the grid~(\ref{eq:Tn_max}), respectively. The
elements of the column matrix $\left [ f \right ]$ are given by
either $f(x_k), \ k=1,\ldots,N$ or $f(\tilde x_k), \
k=0,\ldots,N$. The right-hand side of Eqs.~(\ref{eq:Beqn_a},
\ref{eq:Beqn_b}) and~(\ref{eq:BTeqn_a}, \ref{eq:BTeqn_b}) give the
values of the integral~$\int_{-1}^x \ f(t) \, {\rm d}t$ and the
derivative~$f'(x)$ at the corresponding grid points, respectively.
The actual values of the matrix elements $S_{ij}$ and $D_{ij}$ are
readily available from Eqs.~(\ref{eq:f_derivative_a},
\ref{eq:f_int_a}), while the elements of the matrices $\tilde S$
and $\tilde D$ can be derived using Eqs.~(\ref{eq:f_derivative_b},
\ref{eq:f_int_b}).

\section{Initial value problem}
\label{sec:IVP}

El-gendi~\cite{ref:El-gendi} has extensively shown how Chebyshev
expansions can be used to solve linear integral equations,
integro-differential equations, and ordinary differential
equations on the grid~(\ref{eq:Tn_max}) associated with the (N+1)
extrema of the Chebyshev polynomial of degree~$N$. Also, Delves
and Mohamed have shown~\cite{ref:Delves} that El-gendi's method
represents a modification of the Nystrom scheme when applied to
solving Fredholm integral equations of the second kind. To
summarize these results, we consider first the initial value
problem corresponding to the second-order differential equation
\begin{eqnarray}
   y''(x) \ + \ p(x) \, y'(x) \ + \ q(x) \, y(x) \ = \ r(x)
   \>,
\label{eq:ode2}
\end{eqnarray}
with the initial conditions
\begin{eqnarray}
   y(-1) \ = \ y_0 \>,
   \quad
   y'(-1) \ = \ y'_0
   \>.
\label{eq:initial}
\end{eqnarray}
It is convenient to replace Eqs.(\ref{eq:ode2})
and~(\ref{eq:initial}) by an integral equation, obtained by
integrating twice Eq.~(\ref{eq:ode2}) and using the initial
conditions~(\ref{eq:initial}) to choose the lower bounds of the
integrals.
Equations~(\ref{eq:ode2}) and~(\ref{eq:initial}) reduce to the
integral equation in~$y(x)$
\begin{eqnarray}
   &&
   y(x)
   - y_0
   - (x+1) \Bigl [ y'_0 + p(-1) y_0 \Bigr ]
   +
   \int_{-1}^x  p(x') y(x') {\rm d}x'
   \nonumber \\ &&
   +
   \int_{-1}^x  \int_{-1}^{x'}
        \Bigl [ q(x'') - p'(x'') \Bigr ] y(x'') {\rm d}x'' {\rm d}x'
\label{eq:int_ic}
   \\ && \nonumber
   =
   \int_{-1}^x  \int_{-1}^{x'} r(x'') {\rm d}x'' {\rm d}x'
   \>,
\end{eqnarray}
which is very similar to a Volterra equation of the second kind.
Using the techniques developed in the previous section to
calculate integrals, the integral equation
can be transformed into the linear system of equations
\begin{equation}
   A \ \left [ f \right ] \ = \ C
   \>,
\label{eq:sislin}
\end{equation}
with matrices $A$ and $C$ given as
\begin{eqnarray*}
   A_{i \, j} & = &
   \delta_{i \, j}
   \ + \
   \tilde S_{i \, j} \, p(x_j)
   \ + \
   [\tilde S^2]_{i \, j} \,
                        \Bigl [ q(x_j) - p'(x_j) \Bigr ]
   \>,
   \nonumber \\
   C_i & = & g(x_i) \>,
   \qquad \qquad \qquad
   i,j = 0,1,\ldots,N
   \>.
\end{eqnarray*}
Here the function $g(x)$ is defined by
\begin{eqnarray*}
   g(x) & = &
   \ y_0 \ + \ (x+1) \, \Bigl [ y'_0 \, + \, p(-1) y_0 \Bigr ]
   \nonumber \\ &&
   \ + \ \int_{-1}^x \int_{-1}^{x'} r(x'') \ {\rm d}x'' \ {\rm d}x'
   \>.
\end{eqnarray*}
As a special case we can address the case of the integro-differential
equation:
\begin{equation}
   y''(x) \, + \, p(x) \, y'(x) \, + \, q(x) \, y(x) \, = \,
   \int_{-1}^x \ K(x,t) \, y(t) \, {\rm d}t
   \>,
\label{eq:eqn_dif_int}
\end{equation}
with the initial conditions~(\ref{eq:initial}).
We define the matrix $L$ by
\[
   L_{ij}
   \ = \
   \tilde S_{ij} \, K(x_i, x_j) \>, \quad i,j = 0,1,\ldots,N
   \>.
\]
Then, the solution of the integro-differential
Eq.~(\ref{eq:eqn_dif_int}) subject to the initial
values~(\ref{eq:initial}) can be obtained by solving the system of
$N$ linear equations (\ref{eq:sislin}), where the matrices $A$ and
$C$ are now given by:
\begin{eqnarray*}
   A_{i \, j} & = &
   \delta_{i \, j}
   \ + \
   \tilde S_{i \, j} \, p(x_j)
   \nonumber \\ &&
   \ + \
   [\tilde S^2]_{i \, j} \,
                        \Bigl [ q(x_j)  \, - \, p'(x_j) \Bigr ]
   \ - \
   [\tilde S^2 \, L]_{i \, j}
   \>,
   \nonumber \\
   C_i & = &
   \ y_0 \ + \ (x_i+1) \, \Bigl [ y'_0 \, + \, p(-1) y_0 \Bigr ]
   \>,
\end{eqnarray*}
with $i,j = 0,1,\ldots,N$.

We will illustrate the above method using an example related to
recent calculations of scattering effects in large N expansion and
Schwinger-Dyson equation applications to dynamics in
quantum mechanics~\cite{ref:MDC1
} and quantum field theory~\cite{ref:MDC4}, and compare with
results obtained using traditional finite-difference methods.
Without going into the details of those calculations, it suffices
to say that the crucial step is solving an integral equation of
the form
\begin{eqnarray}
   G(t,t') & = & G_{0}(t,t')
   - 2 \int_{0}^{t}
                 \real \{ Q(t,t'') \} G(t'',t') {\rm d}t''
   \nonumber \\ && \quad
   + 2 \int_{0}^{t'}
                 Q(t,t'') \real \{ G(t'',t') \} {\rm d}t''
   \>,
\label{eq:DQbig0}
\end{eqnarray}
for $G(t,t')$ at positive $t$ and $t'$. Here, $G(t,t')$,
$G_0(t,t')$, and $Q(t,t')$ are complex functions, and the symbols
$\real$ and $\imag$ denote the real and imaginary part,
respectively. In quantum physics applications, the unknown
function $G(t,t')$ plays the role of the two-point Green function
in the Schwinger-Keldysh closed time path
formalism~\cite{ref:CTP}, and obeys the symmetry
\begin{equation}
   G(t,t') \ = \ - \ \overline{G^{}(t',t)} \>,
\label{eq:Dbig_symm}
\end{equation}
where by $\overline{G^{}(t,t')}$ 
we denote the complex conjugate of $G(t,t')$. Therefore the
computation can be restricted to the domain $t' \leq t$.

By separating the real and the imaginary parts of $G(t,t')$,
Eq.~(\ref{eq:DQbig0}) is equivalent to the system of integral
equations
\begin{eqnarray}
   \real \{ G(t,t') \}
   & = &
   \real \{ G_{0}(t,t') \}
\label{eq:ReDQbig0}
   \\ \nonumber  &&
   - 2 \int_{t'}^{t}
                           \real \{ Q(t,t'') \}
                           \real \{ G(t'',t') \} \ {\rm d}t''
   \>,
   \\
   \imag \{ G(t,t') \}
   & = &
   \imag \{ G_{0}(t,t') \}
\label{eq:ImDQbig0}
   \\ \nonumber &&
   - 2 \int_{0}^{t}
                           \real \{ Q(t,t'') \}
                           \imag \{ G(t'',t') \} \ {\rm d}t''
   \\ \nonumber &&
   + 2 \int_{0}^{t'}
                            \imag \{ Q(t,t'') \}
                            \real \{ G(t'',t') \} \ {\rm d}t''
   \>.
\end{eqnarray}
The first equation can be solved for the real part of $G(t,t')$,
and the solution will be used to find $\imag \{ G(t,t') \}$ from
the second equation. This also shows that whatever errors we make
in computing $\real \{ G(t,t') \}$ will worsen the accuracy of the
$\imag \{ G(t,t') \}$ calculation, and thus, $\imag \{ G(t,t') \}$
is a priori more difficult to obtain.

The finite-difference correspondent of Eq.~(\ref{eq:DQbig0}) is
given as
\begin{eqnarray}
   G(i,j) & = & G_0(i,j)
   - 2 \sum_{k=1}^{i-1} \ e_k
                 \real \{ Q(i,k) \} G(k,j)
   \nonumber \\ && \quad
   + 2 \sum_{k=1}^{j-1} e_k
                 Q(i,k) \real \{ G(k,j) \}
   \>,
\label{eq:DQbig0_fd}
\end{eqnarray}
where $e_k$ are the integration weights corresponding to the
various integration methods on the grid. For instance, for the
trapezoidal method, $e_k$ is equal to 1 everywhere except at the
end points, where the weight is 1/2. Note that in deriving
Eq.~(\ref{eq:DQbig0_fd}), we have used the anti-symmetry of the
real part of $G(t,t')$
which gives $\real \{ G(t,t) \} = 0$.

Correspondingly, when using the Chebyshev-expansion with the
grid~(\ref{eq:Tn_max}), the equivalent equation
that needs to be solved is
\begin{eqnarray*}
   G_0(i,j) & = &
   G(i,j)
   \ + \ 2 \sum_{k=0}^{N} \tilde S_{ik}
                 \real \{ Q(i,k) \} G(k,j)
   \nonumber \\ && \quad
   \ - \ 2 \sum_{k=0}^{N} \tilde S_{jk}
                 Q(i,k) \real \{ G(k,j) \}
   \>.
\end{eqnarray*}
In this case the unknown values of $G(t,t')$ on the grid are
obtained as the solution of a system of linear equations.
Moreover, the Chebyshev-expansion approach has the characteristics
of a global method, one obtaining the values of the unknown
function $D(i,j)$ all at once, rather than stepping out the
solution.

\begin{figure}[h!]
   \centering
   \includegraphics[width=2.8in]{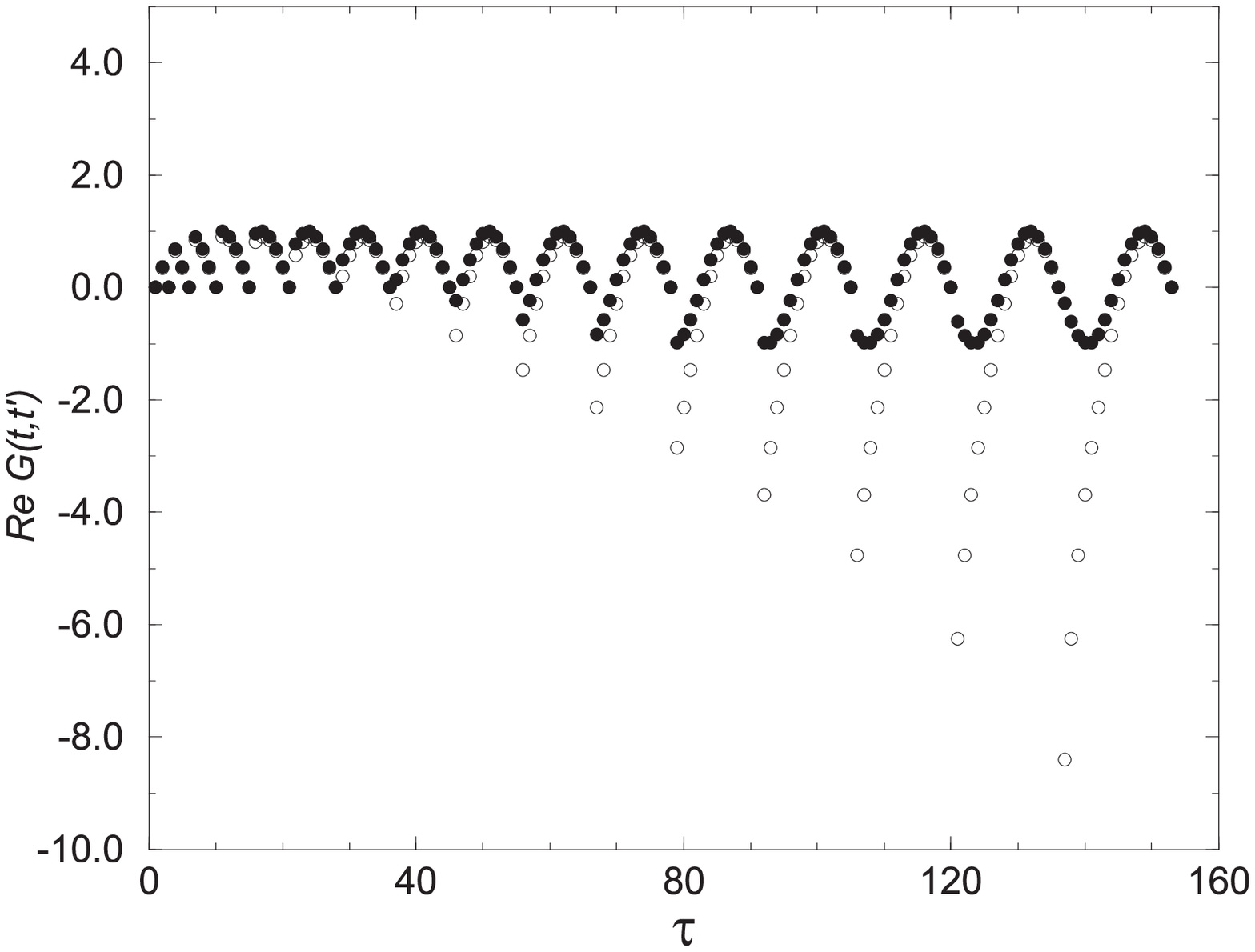}
   \caption{$\real \{ G(t,t') \}$ :
            Chebyshev/exact result (filled) versus the finite-difference result
            corresponding to the trapezoidal method (empty).}
   \label{fig:real_dmat}
\end{figure}
\begin{figure}[h!]
   \centering
   \includegraphics[width=2.8in]{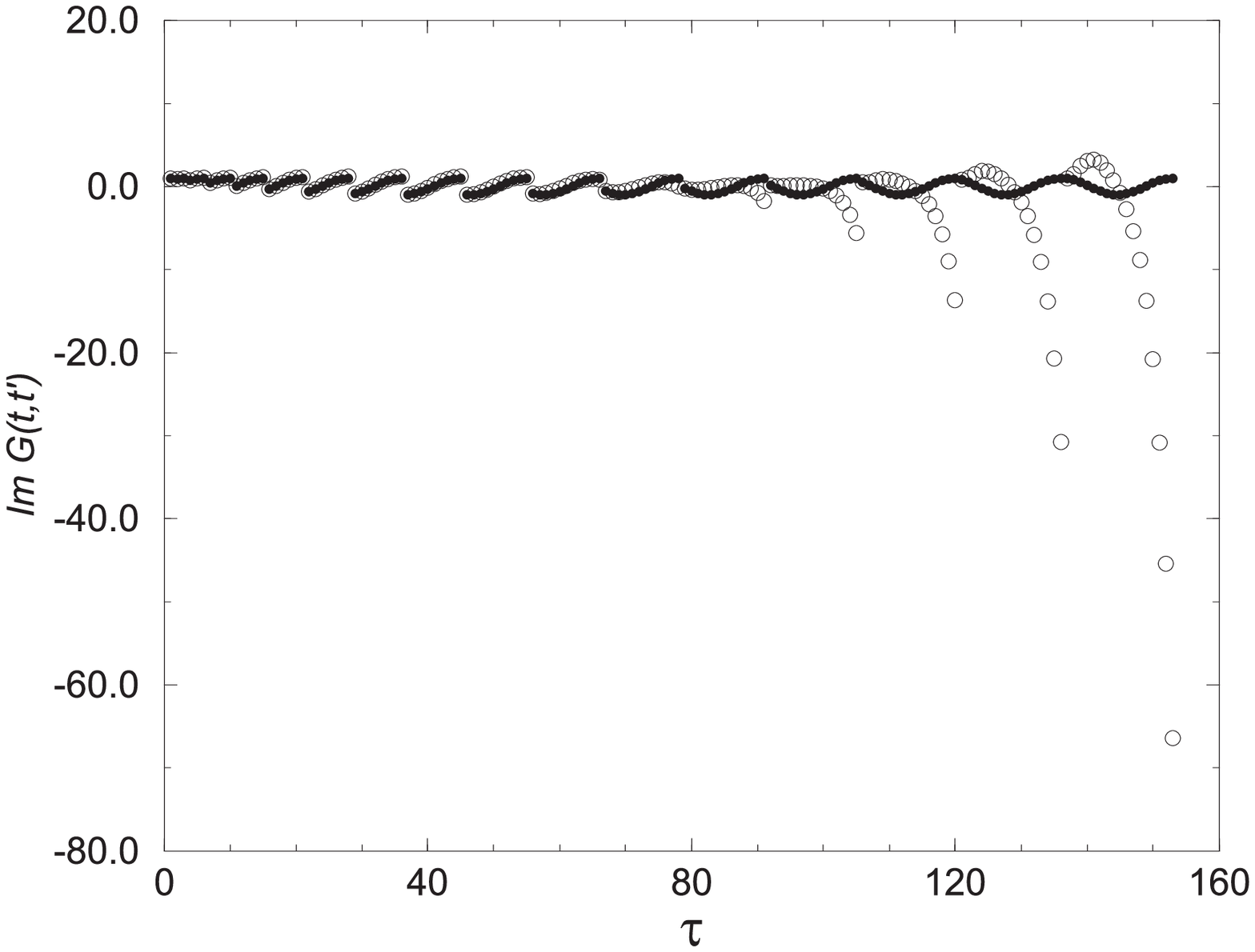}
   \caption{$\imag \{ G(t,t') \}$ :
            Chebyshev/exact result (filled) versus the finite-difference result
            corresponding to the trapezoidal method (empty).}
   \label{fig:imag_dmat}
\end{figure}
In Figs.~\ref{fig:real_dmat} and~\ref{fig:imag_dmat} we compare
the exact result and the finite-difference result corresponding
to the trapezoidal method for a case when the problem has a
closed-form solution. We choose
\begin{eqnarray*}
   Q(t,t') & = & - \sin(t-t') + {\rm i} \cos(t-t')
   \>, \\
   G_0(t,t') & = & (t-t') \cos(t-t')
   \nonumber \\ &&
                   + {\rm i} [ \cos(t-t') - (t+t') \sin(t-t') ]
   \>, \\
   G(t,t') & = & \sin(t-t') + {\rm i} \cos(t-t')
   \>.
\end{eqnarray*}
As we are interested only in the values of $G(t,t')$ for $t' \leq
t$, we depict the real and imaginary parts of $G(t,t')$ as a
function of the band index
$\tau = i(i-1)/2 + j$, with $j\leq i$, used to store the lower
half of the matrix. Given the domain $0 \leq t \leq 6$ and the
same number of grid points ($N$=16), the result obtained using the
Chebyshev expansion approach can not be {\em visually}
distinguished
from the exact result, i.e. the absolute value of the error at
each grid point is less than $10^{-5}$.
As expected we also see that the errors made on $\imag \{ G(t,t')
\}$ by using the finite-difference method are a lot worse than the
errors on $\real \{ G(t,t') \}$. As pointed out before, this is
due to the fact that the equation for $\real \{ G(t,t') \}$ is
independent of any prior knowledge of $\imag \{ G(t,t') \}$ while
we determine $\imag \{ G(t,t') \}$ based on the approximation of
$\real \{ G(t,t') \}$.

\section{Boundary value problem}
\label{sec:BVP}

In principle, the course of action taken in the previous section,
namely converting a differential equation into an integral
equation, also works in the context of a boundary value problem.
Let us consider the second-order ordinary differential equation
\begin{eqnarray}
   y''(x) \ + \ f[y'(x),y(x),x] \ = \ 0
   \>,
   \quad x \in [a,b]
   \>,
\label{eq:ode2_nlin}
\end{eqnarray}
with the boundary conditions
\begin{eqnarray}
   g[y(a),y'(a)] = c_a
   \>,
   \quad
   h[y(b),y'(b)] = c_b
   \>.
\label{eq:bv_cond}
\end{eqnarray}
No restriction on the actual form of the function
$f[y'(x),y(x),x]$ is implied, so 
both linear and nonlinear equations are included.

We integrate Eq.~(\ref{eq:ode2_nlin}) to obtain
\begin{eqnarray}
   y'(x) - y'(a) + \int_a^x f[y'(x'),y(x'),x'] \, {\rm d} x'
   \ = \ 0
   \>.
\label{eq:ode2_yp}
\end{eqnarray}
A second integration gives
\begin{eqnarray}
   &&
   y(x) - y(a) - (x-a) y'(a)
   \nonumber \\ &&
   + \int_a^x \
        \int_a^{x'} f[y'(x''),y(x''),x''] \, {\rm d} x'' \, {\rm d}x'
   \ = \ 0
   \>.
\label{eq:ode2_y}
\end{eqnarray}
The last equation is equivalent to Eq.~(\ref{eq:int_ic}). However,
for an initial value problem, the values of $y(a)$ and $y'(a)$ are
readily available. In order to introduce the boundary conditions
for a boundary value problem, we must consider first a separate
system of equations for $y(a)$, $y(b)$, $y'(a)$ and $y'(b)$, which
is obtained by specializing Eqs.~(\ref{eq:ode2_yp})
and~(\ref{eq:ode2_y}) for $x=b$, together with the boundary
conditions given in (\ref{eq:bv_cond}). Then one can proceed with
solving Eq.~(\ref{eq:ode2_y}) using the techniques presented in
the previous section. For instance, the Dirichlet problem
\begin{equation}
   y(a) \ = \ y(b) \ = \ 0 \>,
\end{equation}
leads to the integral equation
\begin{eqnarray}
   &&
   y(x)
   + \left \{ \frac{x - a}{b - a} \int_a^b + \int_a^x \right \} \
   \\ \nonumber && \times
        \int_a^{x'} f[y'(x''),y(x''),x''] \, {\rm d} x'' \, {\rm d}x'
   \ = \ 0
   \>.
\end{eqnarray}
Note, that one can also double the number of unknowns and solve
Eqs.~(\ref{eq:ode2_yp}) and~(\ref{eq:ode2_y}) simultaneously for
$y(x)$ and $y'(x)$.

In this section however, we will discuss boundary value problems
from the perspective of a perturbative approach, where we start
with an initial guess of the solution $y_0$ that satisfies the
boundary conditions of the problem, and write $y = y_0 +
\epsilon$, with $\epsilon$ being a variation obeying null boundary
conditions. We then solve for the perturbation $\epsilon$ such
that the boundary values remain unchanged. This approach allows us
to treat linear and nonlinear problems on the same footing, and
avoids the additional complications regarding boundary conditions.

We assume that $y_0(x)$ is an approximation of the solution $y(x)$
satisfying the boundary conditions (\ref{eq:bv_cond}). Then we can
write
\[
   y(x) \ = \ y_0(x) \ + \ \epsilon(x)
   \>,
\]
where the variation $\epsilon(x)$ satisfies the boundary conditions
\[
   g[\epsilon(a),\epsilon'(a)] = 0
   \>,
   \quad
   h[\epsilon(b),\epsilon'(b)] = 0
   \>.
\]
We now use the Taylor expansion of $f[y'(x),y(x),x]$ about
$y(x)=y_0(x)$ and keep only the linear terms in $\epsilon(x)$ and
$\epsilon'(x)$ to obtain an equation for the variation
$\epsilon(x)$
\begin{eqnarray}
   \lefteqn{
   \epsilon''(x)
   \, + \,
   \left . \frac{\partial f[y'(x),y(x),x]}{\partial y'(x)}
   \right |_{y(x)=y_0(x)} \epsilon'(x)
   }
   \nonumber \\ &&
   \, + \,
   \left . \frac{\partial f[y'(x),y(x),x]}{\partial y(x)}
   \right |_{y(x)=y_0(x)} \epsilon(x)
   \nonumber \\ &&
   \, = \,
   \, - \, y_0''(x) \, - \, f[y_0'(x),y_0(x),x]
   \>.
\label{eq:eps_eqn}
\end{eqnarray}
Equation~(\ref{eq:eps_eqn}) is of the general form (\ref{eq:ode2})
\begin{displaymath}
   \epsilon''(x) \ + \ p(x) \, \epsilon'(x) \ + \ q(x) \, \epsilon(x) \ = \ r(x)
   \> ,
\end{displaymath}
with
\begin{eqnarray*}
   p(x)
   & = &
   \left . \frac{\partial f[y'(x),y(x),x]}{\partial y'(x)}
   \right |_{y(x)=y_0(x)}
   \>,
   \\
   q(x)
   & = &
   \left . \frac{\partial f[y'(x),y(x),x]}{\partial y(x)}
   \right |_{y(x)=y_0(x)}
   \>,
   \\
   r(x)
   & = &
   - \ y_0''(x) \ - \ f[y_0'(x),y_0(x),x]
   \>.
\end{eqnarray*}
Using the Chebyshev representation of the derivatives,
Eqs.~(\ref{eq:f_derivative_a}, \ref{eq:f_derivative_b}), and
depending on the grid used, we solve a system of linear equations
(\ref{eq:sislin}) for the perturbation function~$\epsilon(x)$.
The elements of the matrices $A$ and $C$ are given as
\begin{eqnarray*}
   A_{ij} & = &
   [ D^2 ]_{i \, j}
   \ + \
   p(x_i) \, D_{i \, j}
   \ + \
   q(x_i) \, \delta_{i \, j} \>,
   \\
   C_i & = & r(x_i) \>,
   \quad  i,j = 1,2,\ldots,N
   \>,
\end{eqnarray*}
for the grid~(\ref{eq:Tn_zeros}), and
\begin{eqnarray*}
   A_{ij} & = &
   [ \tilde D^2 ]_{i \, j}
   \ + \
   p(\tilde x_i) \, \tilde D_{i \, j}
   \ + \
   q(\tilde x_i) \, \delta_{i \, j} \>,
   \\
   C_i & = & r(\tilde x_i) \>,
   \quad  i,j = 1,\ldots,N-1
   \>,
\end{eqnarray*}
for the grid~(\ref{eq:Tn_max}).

The iterative numerical procedure is straightforward:
Starting out with an initial guess $y_0(x)$
we solve Eq.~(\ref{eq:eps_eqn}) for the variation $\epsilon(x)$;
then we calculate the new approximation
of the solution
\begin{equation}
   y_0^{\rm new}
   \ = \
   y_0^{\rm old} \ + \ \epsilon(x)
   \>,
\end{equation}
and repeat the procedure until the difference $\epsilon (x)$ gets
smaller than a certain $\varepsilon$ for all $x$ at the grid points.

\begin{figure}[h!]
   \centering
   \includegraphics[width=2.8in]{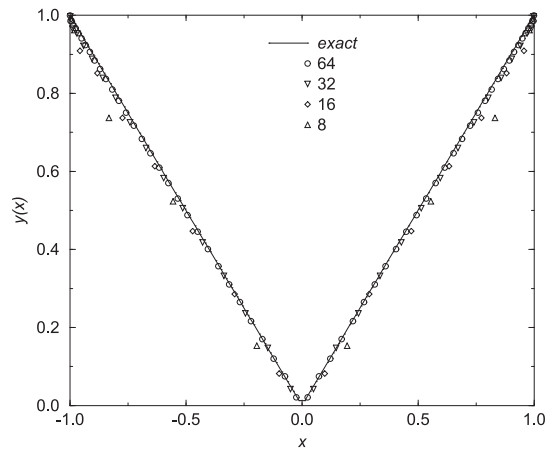}
   \caption{Numerical solutions of
Eq.~(\ref{eq:modul}) obtained for different values of $N$, using
the Chebyshev expansion approach; we chose $y_0(x)=x^2$, for $-1
\leq x \leq 1$}
   \label{fig:fig_abs}
\end{figure}

It is interesting to notice that this approach can work even if
the solution is not differentiable at every point of the interval
where it is defined (Gibbs phenomenon), provided that the lateral
derivatives are finite. As an example, let  us consider the case
of the equation
\begin{equation}
   x \, y'(x) \ - \ y(x) \ = \ 0 \>,
\label{eq:modul}
\end{equation}
which has the solution $y(x) = |x|$. In Fig.~\ref{fig:fig_abs} we
compare the numerical solutions for different values of $N$ on the
interval $[-1,1]$. We see that for $N=64$ the numerical solution
can not be visually discerned from the exact solution.
Eq.~(\ref{eq:modul}) is a good example of a situation when it is
desirable to use an even, rather than an odd, number of grid
points, in order to avoid any direct calculation at the place
where the first derivative $y'(x)$ is not continuous.

\begin{figure}[h!]
   \centering
   \includegraphics[width=2.8in]{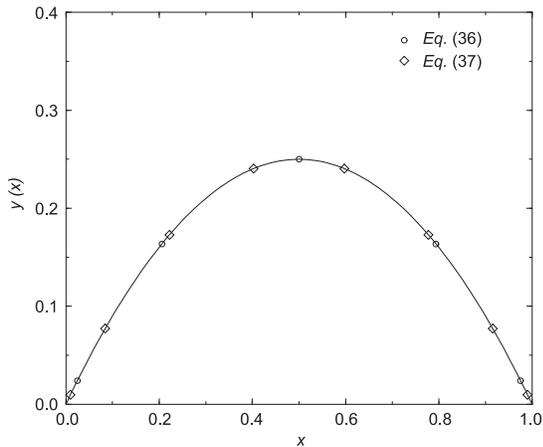}
   \caption{Chebyshev/exact solution of
Eqs.~(\ref{eq:john_1}) and (\ref{eq:john_3}) versus numerical
solutions obtained using the Chebyshev expansion approach.}
   \label{fig:baxley}
\end{figure}

We apply the perturbation approach outlined above to a couple of
singular, nonlinear second-order boundary value problems arising
in fluid dynamics.
The first example~\cite{ref:john_1a
}
        \begin{equation}
           y''(x) \ + \ \frac{\phi(x)}{y^{\lambda}(x)} \ = \ 0 \>,
           \quad \lambda > 0
           \>,
        \label{eq:john_1}
        \end{equation}
gives the Emden-Fowler equation when $\lambda$ is negative.
In order to solve Eq.~(\ref{eq:john_1}), we introduce
the variation $\epsilon(x)$ as a solution of the equation
        \[
           \epsilon''(x)
           \ - \
           \lambda \, \frac{\phi(x)}{y_0^{\lambda+1}(x)} \ \epsilon(x)
           \ = \
           - \ \left \lbrace
                     y_0''(x) + \frac{\phi(x)}{y_0^{\lambda}(x)}
               \right \rbrace
           \>.
        \]
The second example we consider is similar to a particular reduction
of the Navier-Stokes equations~\cite{ref:john_2}
        \begin{equation}
           y''(x) \ - \ \frac{\phi(x)}{y^2(x)} \ y'(x) \ = \ 0
           \>.
        \label{eq:john_3}
        \end{equation}
        In this case,
        the variation $\epsilon(x)$ is a solution of the equation
        \begin{eqnarray*}
           &&
           \epsilon''(x)
           \ - \
           \frac{\phi(x)}{y_0^2(x)} \, \epsilon'(x)
           \ + \
           2 \frac{\phi(x)}{y_0^3(x)} \, y_0'(x) \ \epsilon(x)
           \\ &&
           \ = \
           - \ \left \lbrace
                     y_0''(x) - \frac{\phi(x)}{y_0^2(x)} \, y'(x)
               \right \rbrace
           \>.
        \end{eqnarray*}
In both cases we are seeking solutions $y(x)$ on the interval $[0,1]$,
corresponding to the boundary conditions
\begin{equation}
   y(0) \ = \ y(1) \ = \ 0 \>.
\label{eq:bv_john}
\end{equation}
Then, we choose $y_0(x) = \sin(\pi x)$ as our initial approximation of the solution.
Given the boundary values~(\ref{eq:bv_john}), we see that
the function~$f[y'(x),y(x),x]$ exhibits singularities at both ends
of the interval $[0,1]$. However, since the
variation~$\epsilon(x)$ satisfies null boundary conditions, we
avoid the calculation of any of the coefficients at the singular
points no matter which of the grids~(\ref{eq:Tn_zeros},
\ref{eq:Tn_max}) we choose. We consider the case when the above
problems have the closed-form solution $y(x) = x(1 - x)$, with
$\lambda = 1/2$ in Eq.~(\ref{eq:john_1}). In Fig.~\ref{fig:baxley}
we compare the exact result with the numerical solutions obtained
using the Chebyshev expansion corresponding to the
grid~(\ref{eq:Tn_zeros}).

\begin{figure}[h!]
   \centering
   \includegraphics[width=2.8in]{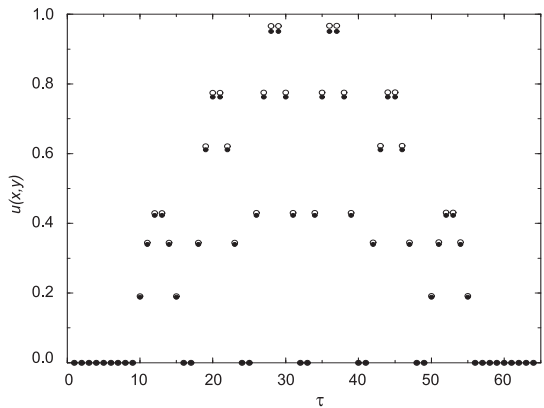}
   \caption{Chebyshev/exact solution (filled) of
Eq.~(\ref{eq:brett}) versus the finite-difference result (empty)
obtained for $N=8$, as a function of the band index $\tau=(i-1)N +
j$.}
   \label{fig:brett}
\end{figure}

The last example we consider arises in the study of ocean currents,
specifically the mathematical explanation of the formation of
currents like the Gulf Stream. Then, one has to solve a partial differential
equation of the type
\begin{equation}
   \left [
   \frac{\partial^2}{\partial x^2} \ + \
   \frac{\partial^2}{\partial y^2} \ + \
   a(x,y) \, \frac{\partial}{\partial x} \right ] \, u(x,y) \ = \ g(x,y)
   \>,
\label{eq:brett}
\end{equation}
subject to null boundary conditions. To illustrate how the method
works in two dimensions, we consider the case of a known solution
$u(x,y) = \sin(\pi x)*\sin(\pi y)$, defined on a square domain
$[0,1]\times[0,1]$ with $a(x,y) = 1$, and compare the results
obtained via a Chebyshev expansion versus the results obtained via
a finite-difference technique. We choose the function $u_0(x,y) =
xy(1-x)(1-y)$ as our initial guess. In Fig.~\ref{fig:brett} we
plot the exact result versus the finite-difference result
corresponding to the same number of points ($n_x=n_y=N$=8) for
which the proposed Chebyshev expansion approach is not
distinguishable from the exact result. The number of iterations
necessary to achieve the desired accuracy is very small (typically
one iteration is enough!), while the finite-difference results are
obtained after 88 iterations.
Of course, the grid can be refined by using a larger number of
mesh points. Then, the number of iterations increases linearly
for the finite-difference method, while the number of iterations
necessary when using the Chebyshev expansion stays pretty much
constant. In general, we do not expect that by using the
Chebyshev expansion, we will always be able to obtain the desired
result after only one iteration. However, the number of necessary
iterations is comparably very small and does not depend
dramatically on the number of grid points. This can be a
considerable advantage when we use a large number of grid points
and want to keep the computation time to a minimum.

\section{Conclusions}
\label{sec:Conclusions}

We have presented practical approaches to the numerical solutions
of initial value and second-order boundary value problems defined
on finite domains, based on a spectral method known as El-gendi's
method. The method is quite general and has some special
advantages for certain classes of problems. This method can be
used also as an initial test to scout the character of the
solution. Failure of the Chebyshev expansion method presented here
should tell us that the solution we seek can not be represented as
a polynomial of order~$N$ on the considered domain.

The Chebyshev grids~(\ref{eq:Tn_zeros}) and (\ref{eq:Tn_max})
provide equally robust ways of discretizing a continuous problem,
the grid~(\ref{eq:Tn_zeros}) allowing one to avoid the calculation
of functions at the ends of the interval, when the equations have
singularities at these points. The fact that the proposed grids
are not uniform should not be considered by any means as a
negative aspect of the method, since the grid can be refined as
much as needed.
The numerical solution in between grid points can always be
obtained by interpolation. The Chebyshev grids have the additional
advantage of being optimal for the cubic spline interpolation
scheme~\cite{ref:spline}.

The Chebyshev expansion provides a robust method of computing the
integral and derivative of a non-singular function defined on a
finite domain. For example, if both the solution of an initial
value problem and its derivative are of interest, it is better to
transform the differential equation into an integral equation and
use the values of the function at the grid points to also compute
the value of the derivative at these points.

\begin{figure}[h!]
   \centering
   \includegraphics[width=2.8in]{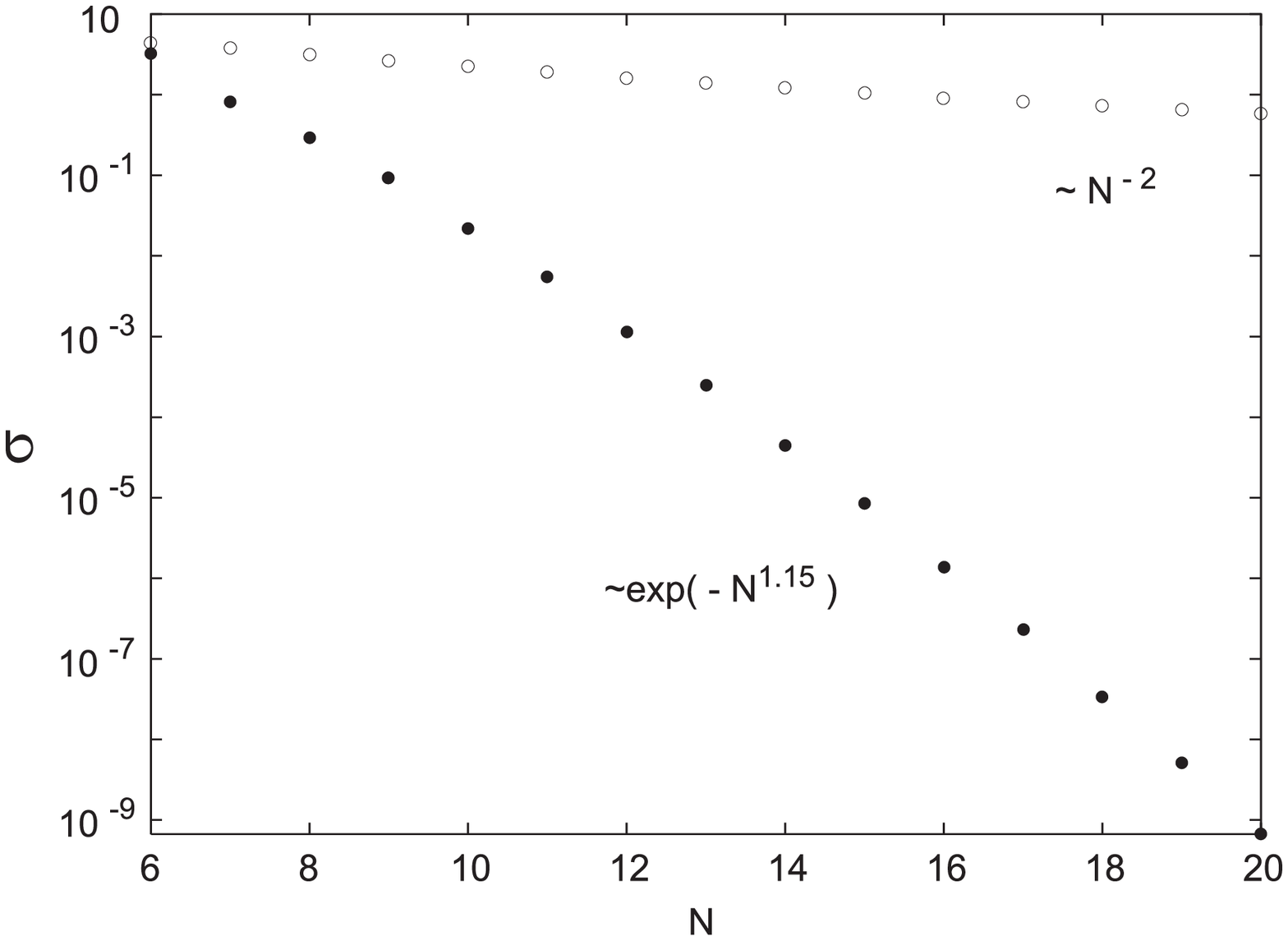}
   \caption{Standard deviation of the approximation
for $\real \{ G(t,t') \}$ as a function of the number of grid
sites: Chebyshev (filled) and finite-difference (empty) results.}
   \label{fig:real_conv}
\end{figure}

\begin{figure}[h!]
   \centering
   \includegraphics[width=2.8in]{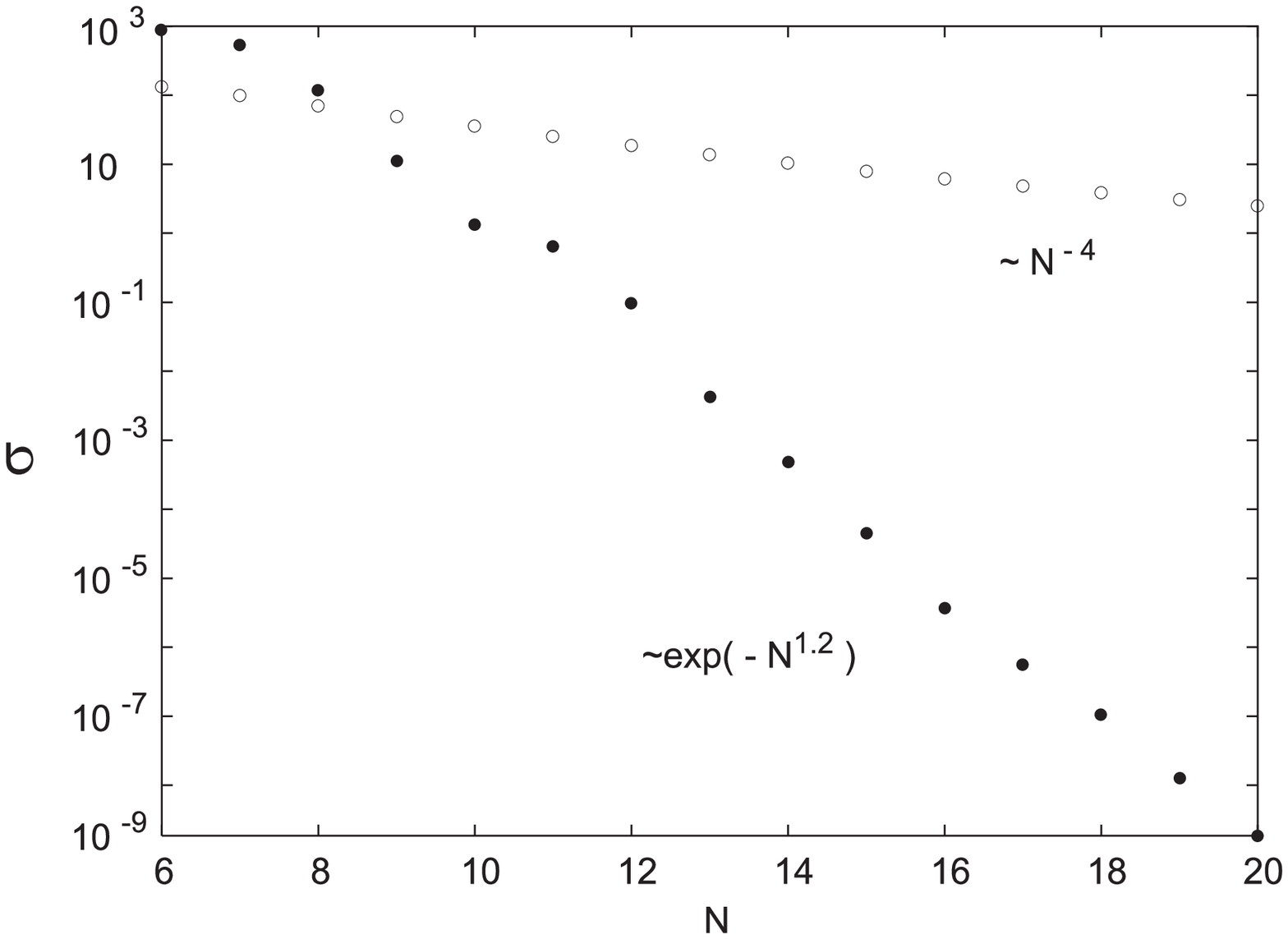}
   \caption{Standard deviation of the approximation
for $\imag \{ G(t,t') \}$ as a function of the number of grid
sites: Chebyshev (filled) and finite-difference (empty) results.}
   \label{fig:imag_conv}
\end{figure}

\begin{figure}[h!]
   \centering
   \includegraphics[width=2.8in]{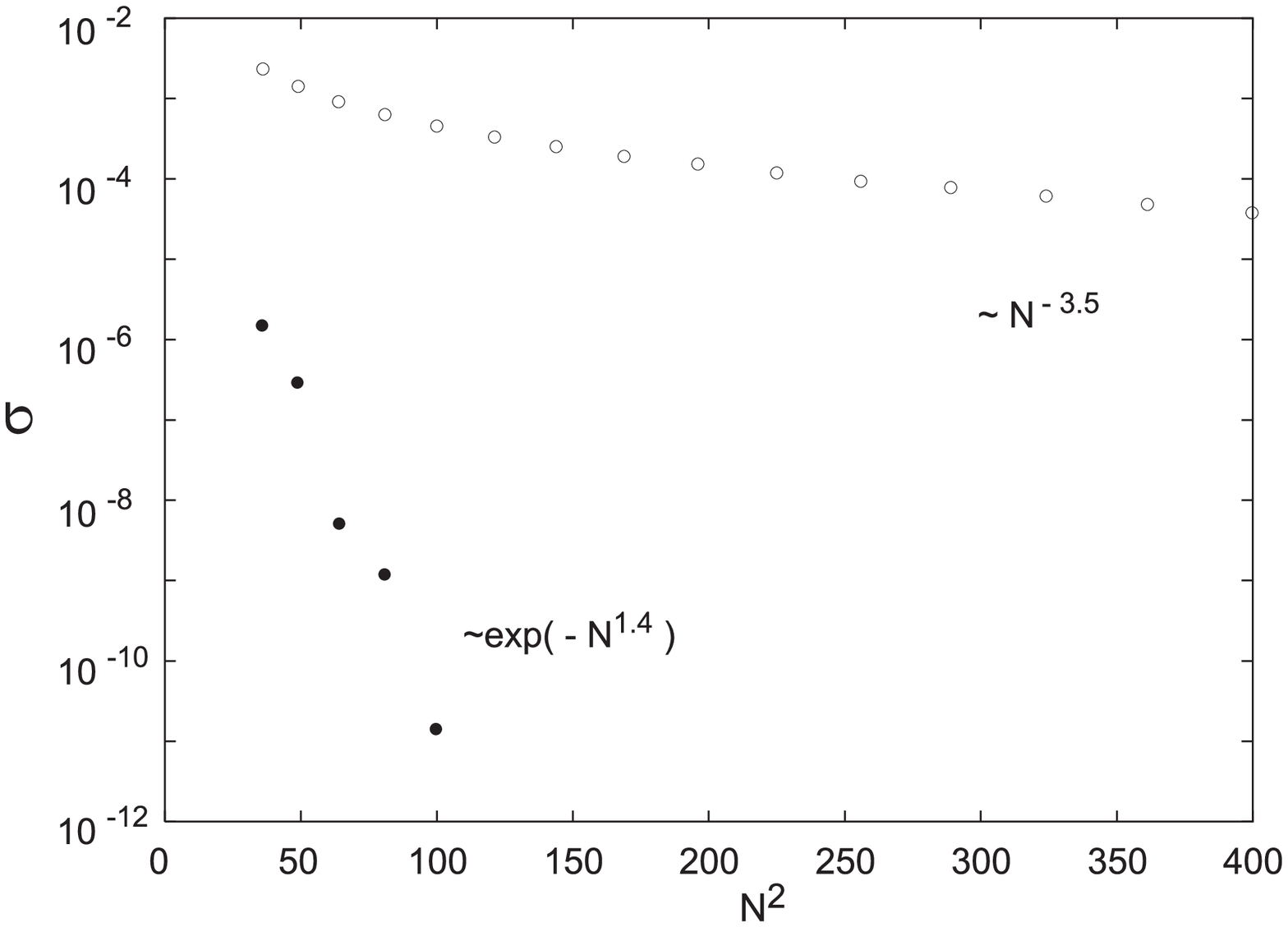}
   \caption{Standard deviation of the approximation
for the solution of Eq.~(\ref{eq:brett}) versus the number of mesh
sites $n_x n_y=N^2$: Chebyshev (filled) and finite-difference
(empty) results.}
   \label{fig:brett_conv}
\end{figure}

\begin{figure}[h]
   \centering
   \includegraphics[width=2.8in]{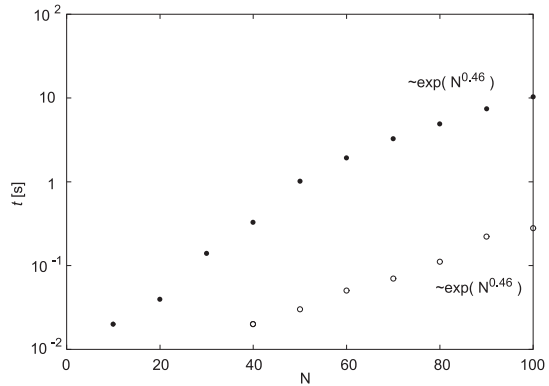}
   \caption{Execution time for obtaining an approximation
for $ G(t,t') $ as a function of the number of grid sites:
Chebyshev (filled) and finite-difference (empty) results.}
   \label{fig:dmat_time}
\end{figure}

\begin{figure}[h]
   \centering
   \includegraphics[width=2.8in]{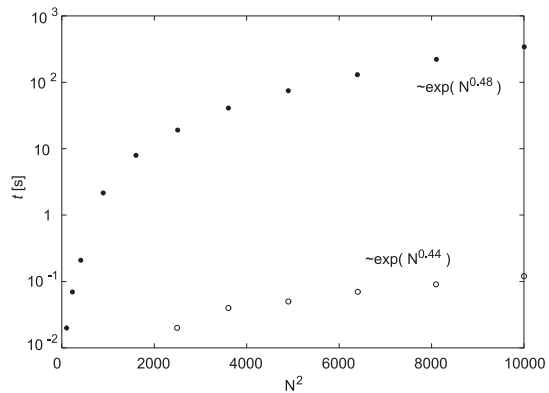}
   \caption{Execution time for obtaining an approximate solution of
Eq.~(\ref{eq:brett}) versus the number of mesh sites $n_x
n_y=N^2$: Chebyshev (filled) and finite-difference (empty)
results.}
   \label{fig:brett_time}
\end{figure}

It is a well-known fact that spectral methods are more expensive
that finite-differences for a given grid size, so in order to
reach some specified accuracy there is always a tradeoff: finite
differences need more points, but are cheaper per point. The goal
is to reach a certain numerical accuracy requirement as
efficiently as possible. Therefore let us discuss here some
computational cost considerations for sample calculations.
We shall comment on two of the calculations presented in this
paper: the calculation for the $G(t,t')$ (see
Eq.~(\ref{eq:DQbig0})) and the solution of Eq.~(\ref{eq:brett}).
In Figs.~\ref{fig:real_conv}, \ref{fig:imag_conv}
and~\ref{fig:brett_conv} we depict the convergence of the
finite-difference and Chebyshev methods for obtaining the
approximate solutions, and we illustrate the elapsed computer time
in Figs.~\ref{fig:dmat_time} and \ref{fig:brett_time}. The error
of the Chebyshev expansion method decays exponentially as a
function of~$N$, while the error of the finite-difference method
can be expressed as a power of~$N$. For both methods, the running
time depends exponentially of~$N$. We conclude that indeed the
execution time required by the spectral method increases faster
with the number of grid points than the finite-difference method.
However, in order to achieve a {\em reasonable} accuracy (e.g.
$\sigma < 10^{-6}$), the Chebyshev method requires only a small
grid, and for this small number of grid points the computer time
is actually modest. All calculations where carried out on a
(rather old) Pentium II 266 MHz PC. There were no additional
numerical algorithms required for performing the finite-difference
calculation, as these involve simple iterations of the initial
guess. Due to the global character of the Chebyshev calculation,
one needs to solve a system of linear equations. Since this is a
sparse system of equations we have employed an iterative
biconjugate gradient method~\cite{ref:nr} for obtaining the
numerical solution. For both problems, the sparsity of the
relevant matrices is~$\sim 2/N$.

Most importantly, we have shown that the Chebyshev expansion is
applicable to efficiently solving complex nonlinear integral
equations of the form encountered in a Schwinger-Dyson approach to
determining the time evolution of the unequal time correlation
functions of non-equilibrium quantum field theory. In this
particular context, spectral methods have made possible for the
first time to carry out complex dynamical calculations at next to
leading order in quantum mechanics and field theory. Our results
will form the basis for future studies of quantum phase
transitions.

\section*{Acknowledgements}

The work of B.M. was supported in part by the U.S. Department of
Energy, Nuclear Physics Division, under contract No.
W-31-109-ENG-38. I.M. would like to thank the hospitality of the
Physics Division of the Argonne National Laboratory, where part of
this project was carried out. The authors gratefully acknowledge
helpful conversations with John Dawson, Fred Cooper, and John
Baxley. The authors would also like to express their gratitude to
the Rocky Mountains Mathematics Consortium, and especially Duane
Porter for providing us with support for attending the RMMC Summer
Conference on Difference Equations and Their Applications,
Laramie, WY, July 13-25, 1997. This was an opportunity for many of
the ideas presented in this paper to take shape, due to fruitful
discussions with others interested in the subject.


\vfill

\end{document}